\documentclass[aps, prd, twocolumn, showpacs, superscriptaddress, groupedaddress, floatfix]{revtex4-2}

\usepackage{graphicx}
\usepackage{yfonts}
\usepackage{float}
\usepackage{amsmath}
\usepackage{amssymb}
\usepackage{amsfonts}
\usepackage{amsthm}
\usepackage{mathtools}
\usepackage{derivative}
\usepackage{dcolumn}
\usepackage{listings}
\usepackage{subfigure, rotating, bm, array}
\usepackage[utf8]{inputenc}
\usepackage[english]{babel}
\usepackage[pagebackref=false, colorlinks=true]{hyperref}
\hypersetup{
linkcolor=blue,     
citecolor=blue,     
urlcolor=blue}      

\usepackage{float}

\begin{document}
\title{The Definition of a Photon Surface in an Invariant Spin Frame}
\author{Dipanjan Dey}
\email{deydipanjan7@gmail.com}
\affiliation{Department of Mathematics and Statistics,
Dalhousie University,
Halifax, Nova Scotia,
Canada B3H 3J5}
\author{Alan A. Coley}
\email{Alan.Coley@dal.ca }
\affiliation{Department of Mathematics and Statistics,
Dalhousie University,
Halifax, Nova Scotia,
Canada B3H 3J5}
\author{Nicholas T. Layden}
\email{nicholas.layden@dal.ca }
\affiliation{Department of Mathematics and Statistics,
Dalhousie University,
Halifax, Nova Scotia,
Canada B3H 3J5}
\date{\today}
\begin{abstract}
This paper defines the photon surface conditions using Cartan scalars within an invariant spin frame, offering a comprehensive description of the local spacetime geometry. By employing this approach, we gain novel insights into the geometry and dynamics of photon surfaces, independent of the global spacetime structure. We first discuss the photon surface conditions in a Petrov type-D spacetime manifold, and then we simplify those conditions assuming the existence of spherical symmetry. Finally, employing the simplified, spherically symmetric photon surface conditions, we explore the dynamics of photon surfaces in static, collapsing Lemaitre-Tolman-Bondi (LTB) spacetimes, and Vaidya spacetimes. Notably, we show that photon surfaces can emerge from the central singularity during the collapse of an inhomogeneous dust cloud modeled by a LTB spacetime. This underscores the significance of our findings in comprehending the potential observational implications of the physics near the ultra-high gravity region.\\
$\textbf{key words}$: Photon surface, Newman-Penrose formalism, Cartan scalars.
\end{abstract}
\maketitle

\section{Introduction}
Recent observations of the shadow of the M87 galaxy center and the Milky Way galaxy center Sgr-A* has triggered a lot of interest in the field of gravitational lensing and the shadow of compact objects \cite{Akiyama:2019fyp}. Specifically, the properties of the shadow and photon rings play a very important role in understanding the nature of the central compact object as well as its surroundings \cite{raji1, AJoshi1, raji2, Patel1, Vir1, Vir2, Vir3, Vir4, Shaikh1, Shaikh2}. The presence of a photon surface around the compact object causes the formation of both the shadow and the photon ring. Consequently, gaining a deeper understanding of the geometry of this photon surface within a generic spacetime manifold can provide fresh insights into the identification of the compact object and the nature of its surroundings.   

A photon surface in general relativity is defined as \cite{Claudel},\\
{\bf Definition:} A photon surface $S$ in a spacetime manifold ($\mathcal{M}, g^{\alpha\beta}$) is an immersed, non-spacelike hypersurface of ($\mathcal{M}, g^{\alpha\beta}$): for every point $p\in S$ and every null vector $l^\mu\in T_p S$, there exists a null geodesic $x^\alpha (\lambda) : (-\epsilon, \epsilon)\to \mathcal{M}$ of $(\mathcal{M}, g^{\alpha\beta})$ such that $\dot{x}^\alpha|_{\lambda = 0}= l^\alpha,~x^\alpha (\lambda)\subset S$.

A photon surface with $SO(3)\times R$ symmetry is known as a photon sphere. A static black hole spacetime serves as an idealized model for real astrophysical black holes. In various astrophysical scenarios, black holes originate from the unhindered gravitational collapse of matter. Once formed, these black holes violently consume the surrounding accreting matter. Consequently, we should not anticipate quiescent black hole dynamics; they should remain dynamically active (classically) as long as there is an accreting matter cloud in their vicinity.

The geometry of the photon surface is intricately linked to the dynamics of the accreting matter surrounding a dynamic black hole. Given that the nature of this accreting matter undergoes violent changes in its dynamics, the photon surface also adapts accordingly. Hence, investigating the dynamic behavior of the photon surface in a generic spacetime is of paramount significance in this context. This study also holds great importance in scenarios where marginally trapped surfaces (MTS) emerge during the unhindered gravitational collapse of a general matter cloud. An MTS is a two-dimensional spacelike hypersurface where the expansion scalar of an outgoing null geodesic congruence becomes zero. This MTS subsequently foliates into a three-dimensional surface known as an apparent horizon. Studying the dynamics of the photon surface in the presence of an apparent horizon would not only provide insights into its behavior during gravitational collapse but also potentially reveal its relationship with the apparent horizon.

Based on the definition of a photon surface given above, one can show the following conditions are equivalent \cite{Claudel}: 
\begin{enumerate}
  \setlength\itemsep{0.005cm}
    \item the non-spacelike surface $S$ is a photon surface;
    \item $K_{\mu\nu}l^\mu l^\nu =0~\forall~ \text{Null}~ l^\mu\in T_pS~\forall p\in S$;
    \item $\sigma_{\mu\nu} = 0$;
    \item every affine null geodesic of $(S, h^{\mu\nu})$ is an affine null geodesic of ($\mathcal{M}, g^{\alpha\beta}$),
\end{enumerate}
where the deformation tensor $K_{\mu\nu} = \nabla_\mu n_\nu$ is defined on the hypersurface $S$ where $n^\mu$ is the normal to $S$ at point $p$. The induced metric on the photon surface is $h_{\mu\nu}$ and the $\sigma_{\mu\nu}$ is the traceless part of $K_{\mu\nu}$. The above conditions are true for all smooth generic spacetime manifolds without any type of symmetry. 

In this paper, we redefine the above photon surface conditions in the corresponding spin frame in an invariant way using the Cartan scalars. Cartan scalars are the projection of the Riemann tensor and the finite number of its derivatives on an invariantly defined null frame. Using the Cartan scalars, one can completely describe the local geometry of a spacetime manifold. In this paper, we explore the geometry of photon surfaces using the Cartan scalars. We first define a canonical null frame where the components of the Riemann tensor take their canonical form and then using the Cartan-Karlhede algorithm, we fix the frame completely. Finally, using the Cartan scalars corresponding to that invariant frame, we define photon surfaces. This approach yields novel insights into the geometry of the photon surface and its dynamics that are not apparent in its previous tensorial definition. Since our spin frame definition of a photon surface is inherently local, similar to the apparent horizon definition, it remains entirely unaffected by the global structure of the spacetime manifold. Here, we show that for some parameters' range, the photon surface can originate from the central singularity formed in the gravitational collapse of the inhomogeneous dust cloud modeled by Lemaitre- Tolman- Bondi (LTB) spacetime, and this property of the photon surface is independent of the external spacetime we consider. These results are very important in the context of the causal structure of the singularity in gravitational collapse.
\begin{center}
\begin{figure}[t]
\includegraphics[scale=0.80]{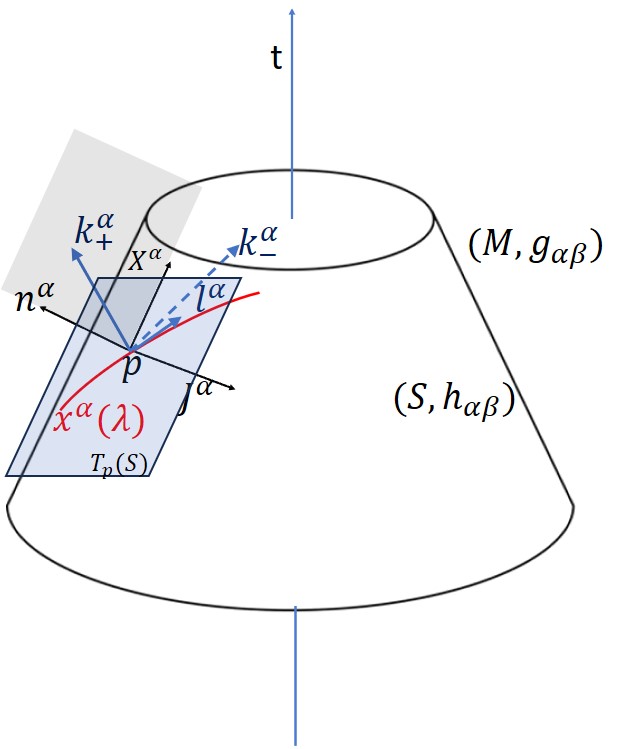}
\caption{ Figure depicts a photon surface $S$, where at point $p$, null vector $l^\alpha \in T_p(S)$ is tangent to the null geodesic (the red curve) on $S$. $X^\alpha \in T_p(S)$ and $J^\alpha \in T_p(S)$ are the timelike and spacelike vectors at $p$, respectively, and $n^\alpha\perp T_p(S)$ is a normal to the surface $S$ at $p$. On the other hand, $k^\alpha_+$ and $k^\alpha_-$ are outgoing and ingoing null vectors, respectively, where they lie on the plane spanned by the $X^\alpha$ and $n^\alpha$.  }
\label{PS}
\end{figure}
\end{center}
The work in this paper is organized in the following way. In section \ref{sec2}, we construct the invariant definition of photon surfaces using Cartan scalars and spin coefficients defined in an invariant null frame. In section \ref{sec3}, we consider a general spherically symmetric static, Petrov type-D solution and define an invariant null frame using the CK algorithm. Then, using the photon surface conditions in terms of Cartan scalars corresponding to that invariant frame, we come up with the well-known coordinate-dependent photon surface condition in a spherically symmetric static spacetime. In the same section, we explore the photon surface geometry and its evolution in collapsing Lemaitre -Tolman-Bondi (LTB) and Vaidya spacetimes using our spin-frame analysis. Finally, in section \ref{sec5}, we discuss the results and possible future work. Throughout the paper, we use a system of units in which the velocity of light and the universal gravitational constant (multiplied by $8\pi$), are both set equal to unity.

\section{Invariant Definition of Photon surface in Spin frame} 
\label{sec2}
In this section, we establish an invariant definition of photon surfaces by employing Cartan scalars and spin coefficients corresponding to an invariantly defined null frame. In particular, we explore the conditions for the existence of photon surfaces in Petrov type-D solutions. Therefore, there exist two principal null directions that one can invariantly define. According to the Cartan-Karlhede (CK) algorithm \cite{cartan-karlhede-alg, kramer}, the initial step involves identifying a null frame in which the Riemann tensor and its derivatives take its type-D canonical form. We refer to a null frame that fulfills this condition as a canonical frame. After identifying the canonical frame, one needs to invariantly fix the frame further by using the CK algorithm. The steps of the CK algorithm are as follows:
\begin{enumerate}
    \item Initialize the differentiation order, denoted as $q$, to 0.
    \item Compute the derivatives of the Riemann tensor up to the $q$th order.
    \item Determine the canonical form of both the Riemann tensor and its derivatives.
    \item Fix the frame as much as possible based on this canonical form. Take note of any remaining frame freedom, identified as the linear isotropy group $\hat{I}_q$. 
    \item Identify $t_q$, the number of independent functions associated with the space-time position in the components of the Riemann tensor and its derivatives in canonical form. 
    \item If the isotropy group and the number of independent functions match those from the previous step, set $p + 1 = q$ and conclude the procedure. If they differ (or if $q = 0$), increment $q$ by 1 and return to step 2.
\end{enumerate}
After using the above-mentioned algorithm, we can invariantly define a null frame at every point in a spacetime manifold and, consequently, we can express any local property of the given spacetime manifold in terms of the Cartan scalars corresponding to that invariant null frame. For a spherically symmetric Petrov type-D solution, it can be shown that a null frame can be invariantly defined up to the spatial spin. For type-D non-spherical solutions, one has to find an invariant way to fix the frame completely by fixing the spatial spin \cite{Nick1, Nick2}. It should be noted that in this paper, when we refer to a null frame as an invariant frame we mean it is invariant up to the spatial spin. 

After establishing an invariant null frame $(k^\alpha_+, k^\alpha_-, m^\alpha, \overline{m}^\alpha)$ at a specific spacetime point $p$, the next step to invariantly define a photon surface involves determining the Lorentz transformation that converts this invariant frame to the null frame $(l^\alpha, p^\alpha, M^\alpha, \overline{M}^\alpha)$, also defined at the same point $p$. Here, $l^\alpha$ denotes the null vector tangent to a three-dimensional hypersurface $S$, and it must satisfy the previously mentioned photon surface condition if and only if the hypersurface $S$ qualifies as a photon surface  (see Fig.~(\ref{PS})). Since $l^\alpha, p^\alpha, M^\alpha, \overline{M}^\alpha, k^\alpha_+, k^\alpha_-, m^\alpha,\overline{m}^\alpha$ all are null vectors, they satisfy the following null conditions: $l^\alpha p_\alpha =-1, l^\alpha M_\alpha= l^\alpha \overline{M}_\alpha =0,~ p^\alpha M_\alpha= p^\alpha \overline{M}_\alpha =0, M^\alpha \overline{M}_\alpha =1$ and $k^\alpha_+ k_{\alpha-} =-1, k^\alpha_+ m_\alpha= k^\alpha_+ \overline{m}_\alpha =0, k^\alpha_- m_\alpha= k^\alpha_- \overline{m}_\alpha =0, m^\alpha \overline{m}_\alpha =1$. In this context, $k^\alpha_+$ and $k^\alpha_-$ denote the outgoing and incoming null bases, respectively, within the null frame $(k^\alpha_+, k^\alpha_-, m^\alpha, \overline{m}^\alpha)$. Going to a special null frame is always advantageous where two of the four null bases possess outgoing and ingoing characteristics. In such a frame, the curvature and its derivatives typically take on their canonical forms. In this context, ``outgoing" and ``ingoing" refer to the positive and negative values of the expansion scalar of the corresponding null vector congruence, respectively  (assuming the absence of trapped surfaces). Now, let's consider that the frame $(k^\alpha_+, k^\alpha_-, m^\alpha,\overline{m}^\alpha)$ is a canonical frame which is defined invariantly by using Cartan-Karlhede (CK) algorithm \cite{cartan-karlhede-alg, kramer}. There can be a combination of the following four Lorentz transformations of the invariant frame $(k^\alpha_+, k^\alpha_-, m^\alpha,\overline{m}^\alpha):$
\begin{eqnarray}
    &&\text{$1.$ Boost in the $(k^\alpha_+, k^\alpha_-)$ plane:}\nonumber \\
&k&^{\prime\alpha}_+ = A k^\alpha_+\, , \,k^{\prime\alpha}_- = \frac{1}{A} k^\alpha_-\, , m^{\prime\alpha}=m^\alpha\, , \overline{m}^{\prime\alpha}=\overline{m}^\alpha \,\label{Lboost} \\
\nonumber\\
&&\text{$2.$ Null rotation making $k^\alpha_-$ fixed:}\nonumber \\
&k&^{\prime\alpha}_- = k^\alpha_-\, ,\,k^{\prime\alpha}_+ = k^\alpha_+ + E \overline{m}^\alpha + \overline{E} m^\alpha + E\overline{E}k^\alpha_- ,\nonumber\\
&m&^{\prime\alpha} = m^\alpha +Ek^\alpha_-\label{Lnullrot1}\\
\nonumber\\
&&\text{$3.$ Null rotation making $k^\alpha_+$ fixed:}\nonumber \\
&k&^{\prime\alpha}_+ = k^\alpha_+\, ,\,k^{\prime\alpha}_- = k^\alpha_- + B \overline{m}^\alpha + \overline{B} m^\alpha + B\overline{B}k^\alpha_+,\nonumber\\
&m&^{\prime\alpha} = m^\alpha +Bk^\alpha_+\label{Lnullrot2}\\
\nonumber\\
&&\text{$4.$ Spatial rotation in the $(m^\alpha, \overline{m}^\alpha)$ plane:}\nonumber \\
&k&^{\prime\alpha}_+ = k^\alpha_+\, ,\,k^{\prime\alpha}_- = k^\alpha_-\, , m^{\prime\alpha}=e^{i\Theta}m^\alpha\, , \overline{m}^{\prime\alpha}=e^{-i\Theta}\overline{m}^\alpha\,\, \label{Lspin},
\end{eqnarray}
where $A$ is the real boost parameter, $E$ and $B$ are complex null rotation parameters, and the $\Theta$ is the real spin parameter and all are functions of all of the spacetime coordinates in general. Now, from the invariant null frame $(k^\alpha_+, k^\alpha_-, m^\alpha,\overline{m}^\alpha)$, we construct the null frame $(l^\alpha, p^\alpha, M^\alpha, \overline{M}^\alpha)$ after four successive Lorentz transformations discussed above. Then the null bases of $(l^\alpha, p^\alpha, M^\alpha, \overline{M}^\alpha)$ can be written as:
\begin{widetext}
\begin{eqnarray}
    l^\mu &=& A k^\mu_+ + \frac{E\Bar{E}}{A} k^\mu_- + \overline{E}m^\mu + E\overline{m}^\mu,\nonumber\\
    p^\mu &=& AB\overline{B} ~k^\mu_+ + \frac{1}{A}\left(1+E\overline{E}B\overline{B}+B\overline{E}+E\overline{B}\right) k^\mu_- + \overline{B}\left(B\overline{E}+1\right)m^\mu + B\left(\overline{B}E + 1\right)\overline{m}^\mu,\nonumber\\
    M^\mu &=& e^{i\Theta}\left(BA~ k^\mu_+ + \frac{E}{A}\left(B\overline{E}+1\right) k^\mu_- +\left(B\overline{E}+1\right)m^\mu + BE~\overline{m}^\mu\right),\nonumber\\
    \overline{M}^\mu &=& e^{-i\Theta}\left(\overline{B}A~ k^\mu_+ + \frac{\overline{E}}{A}\left(\overline{B}E+1\right) k^\mu_- +\overline{B}\overline{E}~m^\mu + \left(\overline{B}E+1\right)~\overline{m}^\mu\right).\nonumber\\
    \label{nullframegen}
\end{eqnarray}
\end{widetext}
Now, if the above-mentioned three-dimensional surface $S$ is a photon surface then the previously mentioned condition $K_{\mu\nu} l^\mu l^\nu = 0$ should hold on the surface. Using the expressions for $M^\mu$ and $\overline{M}^\mu$ in Eq.~(\ref{nullframegen}), we can construct a spacelike normal $n^\mu$ to $T_p(S)$: $n^\mu = \left(\mathcal{N}M^\mu + \frac{1}{2\mathcal{N}}\overline{M}^\mu\right)$, where the normal is orthogonal to $l^\mu$ and $p^\mu$ and $\mathcal{N}$ is a real-valued constant, and consequently, we can write:
\begin{eqnarray}
 K_{\mu\nu} l^\mu l^\nu = 0 \implies l^\mu l^\nu \left(\mathcal{N}\nabla_\mu M_\nu +  \frac{1}{2\mathcal{N}}\nabla_\mu \overline{M}_\nu\right) &=& 0\nonumber\\
 \implies \mathcal{N}M^\nu \Tilde{D}l_\nu +  \frac{1}{2\mathcal{N}}\overline{M}^\nu \Tilde{D}l_\nu = \mathcal{N}\Tilde{\kappa} +  \frac{1}{2\mathcal{N}}\overline{\Tilde{\kappa}} &=& 0, \nonumber\\
\end{eqnarray}
where $\Tilde{D} \equiv l^\mu\nabla_\mu$ and $\Tilde{\kappa} \equiv M^\nu \Tilde{D}l_\nu$ is one of the spin coefficients in the $(l^\alpha, p^\alpha, M^\alpha, \overline{M}^\alpha)$ frame. The above equation is true for all real values of $\mathcal{N}$. Therefore, the above equation implies $\Tilde{\kappa} = \overline{\Tilde{\kappa}} = 0$ which is somewhat obvious since $\Tilde{D}l_\alpha = \left(\Tilde{\epsilon}+\overline{\Tilde{\epsilon}}\right)l_\alpha - \overline{\Tilde{\kappa}} M_\alpha-\Tilde{\kappa}\overline{M}_\alpha$, where the spin coefficient $\Tilde{\epsilon}=\frac12\left(\overline{M}^\alpha~DM_\alpha~-~p^\alpha~Dl_\alpha\right)$. On the photon surface $S$ the $\Tilde{D}l_\alpha = 0$, which implies $\Tilde{\kappa} = \overline{\Tilde{\kappa}} = 0$ and $\left(\Tilde{\epsilon}+\overline{\Tilde{\epsilon}}\right) = 0$. However, the condition $\Tilde{\kappa} = \overline{\Tilde{\kappa}} = 0$ is not an invariantly defined condition for the photon surface, since $(l^\alpha, p^\alpha, M^\alpha, \overline{M}^\alpha)$ is not an invariant frame. For example, if the spacetime is Petrov type-D and the $(k^\alpha_+, k^\alpha_-, m^\alpha,\overline{m}^\alpha)$ null frame is the canonical invariant frame (up to spatial spin) then in that frame the only non-zero component of Weyl tensor ($C_{\mu\nu\alpha\beta}$) is $\Psi_2$, where $\Psi_2 = C_{\mu\nu\alpha\beta} k^\mu_+ m^\nu \overline{m}^\alpha  k^\beta_-$. However, one can show that the components of the Weyl curvature tensor do not have the canonical form in the null frame $(l^\alpha, p^\alpha, M^\alpha, \overline{M}^\alpha)$, in which the Weyl tensor components are: 
\begin{eqnarray}
    \Tilde{\Psi}_0 &=& 6~E^2~e^{2i\Theta}~\Psi_2\, ,\nonumber\\
    \Tilde{\Psi}_1 &=& 3~E\left(1 + 2\overline{B}E\right)~e^{i\Theta}~\Psi_2\, ,\nonumber\\
    \Tilde{\Psi}_2 &=& \left(1 + 6\overline{B}E + 6\overline{B}^2E^2 \right)~\Psi_2\, ,\nonumber\\
    \Tilde{\Psi}_3 &=& 3\overline{B}\left(1+3\overline{B}E + 2\overline{B}^2E^2\right)~e^{-i\Theta}~\Psi_2\, ,\nonumber\\
    \Tilde{\Psi}_4 &=& 6\overline{B}^2\left(1+2\overline{B}E + \overline{B}^2E^2\right)~e^{-2i\Theta}~\Psi_2\, ,
    \label{psi}
\end{eqnarray}
where $\Psi_0 = C_{\mu\nu\alpha\beta} k^\mu_+ m^\nu k^\alpha_+ m^\beta$, $\Psi_1 = C_{\mu\nu\alpha\beta} k^\mu_+ k^\nu_- k^\alpha_+ m^\beta$, $\Psi_3 = C_{\mu\nu\alpha\beta} k^\mu_+ k^\nu_- \overline{m}^\alpha  k^\beta_-$, $\Psi_4 = C_{\mu\nu\alpha\beta} \overline{m}^\mu k^\nu_- \overline{m}^\alpha  k^\beta_-$, and $\tilde{\Psi}_0, \tilde{\Psi}_1, \tilde{\Psi}_2, \tilde{\Psi}_3, \tilde{\Psi}_4$ are the corresponding components of Weyl tensor in the $(l^\alpha, p^\alpha, M^\alpha, \overline{M}^\alpha)$ frame. The above equations (Eq.~(\ref{psi})) show that the components of the Weyl curvature tensor do not have a canonical form in the $(l^\alpha, p^\alpha, M^\alpha, \overline{M}^\alpha)$ frame. To express the photon surface condition in terms of the spin-coefficients of the invariantly defined null frame $(k^\alpha_+, k^\alpha_-, m^\alpha,\overline{m}^\alpha)$, we need to find out how the condition $\Tilde{\kappa} = \overline{\Tilde{\kappa}} = 0$ changes after the composite Lorentz transformations mentioned above. It can be verified that the condition $\Tilde{\kappa} = \overline{\Tilde{\kappa}} = 0$ is transformed to the following condition in the invariant frame:
\begin{widetext}
\begin{eqnarray}
  &A&^2\kappa + AE\left(2\epsilon +\rho\right)+A\overline{E}~\sigma+E^2\left(\pi + 2\alpha\right)
  +E\overline{E}\left(\tau + 2\beta\right) + \frac{E^3}{A}\lambda +\frac{E^2\overline{E}}{A}\left(\mu+2\gamma\right)
  +\frac{E^3\overline{E}}{A^2}\nu\nonumber\\
  &+&~E\left(DA + \frac{E\overline{E}}{A^2}~\Delta A+\frac{\overline{E}}{A}~\delta A +\frac{E}{A}~\overline{\delta}A\right)
  +\left(\overline{E}E~DE+\Delta E+\overline{E}\overline{\delta}E+E\delta E\right)~=~0\, ,
  \label{PScon1}
\end{eqnarray}   
\end{widetext}
where $\pi, \nu, \kappa, \tau, \lambda, \sigma, \alpha, \beta, \mu, \rho, \gamma, \epsilon$ are the spin coefficients in the invariant frame $(k^\alpha_+, k^\alpha_-, m^\alpha,\overline{m}^\alpha)$ and an over-bar on them denotes the complex conjugates of them and $D= k^\alpha_+\nabla_\alpha,~ \Delta= k^\alpha_-\nabla_\alpha,~ \delta= m^\alpha\nabla_\alpha,~ \overline{\delta}= \overline{m}^\alpha\nabla_\alpha$ are the corresponding frame derivatives \cite{kramer}. The above condition for photon surfaces in terms of spin coefficients of the invariant null frame $(k^\alpha_+, k^\alpha_-, m^\alpha,\overline{m}^\alpha)$ is a generic one that is independent of any type of symmetry. Now we can construct a timelike vector $X^\alpha: X^\alpha\in T_p(S)$ and a spacelike vector $J^\alpha: J^\alpha\in T_p(S)$. One can write down the general form of $X^\alpha$ and $J^\alpha$ as:
\begin{widetext}
\begin{eqnarray}
    &&X^\alpha = \mathcal{X}~l^\alpha + \frac{1}{2\mathcal{X}}~p^\alpha\nonumber\\
    &=& \frac{A\left(2\mathcal{X}^2 + B\overline{B}\right)}{2\mathcal{X}} k^\alpha_+ + \frac{\left(1+\overline{B}E+B\overline{E}+E\overline{E}\left(2\mathcal{X}^2+B\overline{B}\right)\right)}{2A\mathcal{X}}k^\alpha_- + \frac{\left(\overline{E}\left(2\mathcal{X}^2 + B\overline{B}\right)+\overline{B}\right)}{2\mathcal{X}}m^\alpha + \frac{\left(E\left(2\mathcal{X}^2 + B\overline{B}\right)+B\right)}{2\mathcal{X}}\overline{m}^\alpha\, ,\nonumber\\
    &&J^\alpha = \mathcal{X}~l^\alpha - \frac{1}{2\mathcal{X}}~p^\alpha\nonumber\\
    &=& \frac{A\left(2\mathcal{X}^2 - B\overline{B}\right)}{2\mathcal{X}} k^\alpha_+ + \frac{\left(E\overline{E}\left(2\mathcal{X}^2-B\overline{B}\right)-\overline{B}E-B\overline{E}-1\right)}{2A\mathcal{X}}k^\alpha_- + \frac{\left(\overline{E}\left(2\mathcal{X}^2 - B\overline{B}\right)-\overline{B}\right)}{2\mathcal{X}}m^\alpha + \frac{\left(E\left(2\mathcal{X}^2 - B\overline{B}\right)-B\right)}{2\mathcal{X}}\overline{m}^\alpha\, ,\nonumber\\
\end{eqnarray}
\end{widetext}
where $\mathcal{X}$ is a real-valued constant. 
Now, let's define an orthonormal frame $(u^\alpha, e^\alpha_{(1)}, e^\alpha_{(2)}, e^\alpha_{(3)})$ corresponding to the invariant frame $(k^\alpha_+, k^\alpha_-, m^\alpha,\overline{m}^\alpha)$ such that one can write: $k^\alpha_+ = \mathcal{U}(u^\alpha + e^\alpha_{(1)}),~ k^\alpha_- = \frac{1}{2\mathcal{U}}(u^\alpha - e^\alpha_{(1)}),~m^\alpha = \frac{1}{\sqrt{2}}(e^\alpha_{(2)} + i e^\alpha_{(3)}),~\overline{m}^\alpha = \frac{1}{\sqrt{2}}(e^\alpha_{(2)} - i e^\alpha_{(3)})$, where $u^\alpha$ is timelike in nature and the other three bases are spacelike, and $\mathcal{U}$ and $\mathcal{E}$ are real-valued constants.
The above expression of the timelike vector $X^\alpha$ illustrates that in a general scenario, it has non-zero components in the directions of all the four bases of the orthonormal frame $(u^\alpha, e^\alpha_{(1)}, e^\alpha_{(2)}, e^\alpha_{(3)})$.

As stated above, the condition for the existence of the photon surface, as described in Eq.~(\ref{PScon1}), is defined in an invariant frame and is independent of any symmetry of the spacetime manifold. Although the formulation may seem somewhat lengthy, it could prove more insightful than the tensorial expression $K_{\mu\nu}l^\mu l^\nu = 0$. Given our understanding of how spin coefficients and curvature components in an invariant null frame behave across different classes of exact solutions to Einstein's equations, we can systematically verify the presence of photon surfaces within diverse spacetime manifolds and this approach allows us to gain a deeper understanding of the geometric and dynamic structure of a photon surface, revealing aspects that may not be readily apparent in the tensorial representation. Moreover, we can use the above equation to redefine the photon surface condition using Cartan scalars. The above condition (Eq.~(\ref{PScon1})) of the photon surface can be written concisely as $f(\mathcal{S}) = \overline{f}(\overline{\mathcal{S}}) = 0$, where $f$ is the function of a set of spin coefficients represented by $\mathcal{S}$. Since this condition is always true on the photon surface, we can define the normal $n^\mu$ at any point on the surface as:
\begin{eqnarray}
    n_\alpha = \nabla_\alpha f(\mathcal{S})|_{f(\mathcal{S})=0}=\nabla_\alpha \overline{f}(\overline{\mathcal{S}})|_{\overline{f}(\overline{\mathcal{S}})=0}.
\end{eqnarray}
Now, since $l^\alpha n_\alpha =0$, using the expression of $l^\alpha$ in Eq.~(\ref{nullframe}), we get:
\begin{eqnarray}
    \left[A~Df(\mathcal{S}) + \frac{E\overline{E}}{A}~ \Delta f(\mathcal{S})+ \overline{E}~\delta f(\mathcal{S}) + E~\overline{\delta}f(\mathcal{S})\right]_{f(\mathcal{S}) = 0} = 0.\nonumber\\
    \label{PScon2}
\end{eqnarray}
The above equation can provide us with the photon surface condition in terms of Cartan scalars defined in the invariant frame $(k^\alpha_+, k^\alpha_-, m^\alpha,\overline{m}^\alpha)$.
\\
\section{Photon surfaces in spherically symmetric solution}
\label{sec3}
In a spherically symmetric scenario, we can always construct at least three spacelike Killing vectors $\mathcal{K}_1, \mathcal{K}_2, \mathcal{K}_3$ such that $\mathcal{L}_{\mathcal{K}_1}g = \mathcal{L}_{\mathcal{K}_2}g = \mathcal{L}_{\mathcal{K}_3}g = 0$, where $\mathcal{L}_{\mathcal{K}_1}$, $\mathcal{L}_{\mathcal{K}_2}$ and $\mathcal{L}_{\mathcal{K}_3}$ are the Lie derivative along the $\mathcal{K}_1$, $ \mathcal{K}_2$, and $ \mathcal{K}_3$ respectively and $g$ is the metric tensor. These three Killing vectors are always orthogonal to one timelike vector and one spacelike vector. Now, let's consider the invariant frame $(k^\alpha_+, k^\alpha_-, m^\alpha,\overline{m}^\alpha)$ is oriented in such a way that $g({\bf u}, {\bf\mathcal{K}_i}) = g({\bf e_{(1)}}, {\bf\mathcal{K}_i}) = 0$, where $i$ stands for $1,2,3$ and 
\begin{eqnarray}
    \mathcal{K}_3 &=& \partial_{(3)}\nonumber\\
    \mathcal{K}_1 &=& -\sin(y^{(3)})~\partial_{(2)}-\cot(y^{(2)})\cos(y^{(3)})~\partial_{(3)}\nonumber\\
    \mathcal{K}_2 &=& \cos(y^{(3)})~\partial_{(2)}-\cot(y^{(2)})\sin(y^{(3)})~\partial_{(3)},
\end{eqnarray}
where the coordinates $y^{(2)}$ and $y^{(3)}$ correspond to the bases $\partial_{(2)}$ and $\partial_{(3)}$, respectively. Among the three Killing vectors, two are linearly independent and span a two-dimensional vector space at point $p\in \mathcal{M}$. Any vector in this two-dimensional spacelike vector space is orthogonal to the bases ${\bf u}$ and ${\bf e_1}$ and, consequently, they are orthogonal to the null vectors ${\bf k_+}$, ${\bf k_-}$. For the spherically symmetric scenarios, the timelike vector $X^\alpha$ tangent to the photon surface $S$ must be orthogonal to a two-dimensional spacelike spherically symmetric hypersurface $\Sigma \subset S$.
Therefore, $X^\alpha$ must not have any components in the null directions $m^\alpha$ and $\overline{m}^\alpha$. Hence, in spherically symmetric spacetimes, $$\left(E\left(2\mathcal{X}^2 + B\overline{B}\right)+B\right) = 0,$$ which implies 
\begin{eqnarray}
    X^\alpha = -\frac{AB}{2\mathcal{X}E} ~k^\alpha_+ - \frac{\mathcal{X}E}{AB}~k^\alpha_-,
\end{eqnarray}
and consequently, the spacelike vector $J^\alpha$ becomes:
\begin{eqnarray}
J^\alpha=\frac{A\left(2\mathcal{X}^2 - B\overline{B}\right)}{2\mathcal{X}} k^\alpha_+ &-& \frac{\mathcal{X}~E\overline{E}\left(2\mathcal{X}^2 - B\overline{B}\right)}{AB\overline{B}}k^\alpha_-\nonumber\\ &+& 2\mathcal{X}\overline{E}m^\alpha + 2\mathcal{X}E\overline{m}^\alpha.   
\end{eqnarray}
For the spherically symmetric scenario, $J^\alpha$ is tangent to the two-dimensional spacelike hypersurface $\Sigma$ at any point $p$, and that surface is orthogonal to the bases $u^\alpha$ and $e^\alpha_{(1)}$. Therefore, $g({\bf u}, {\bf J}) = g({\bf e_{(1)}}, {\bf J}) = 0$, which implies: $B\overline{B} = 2\mathcal{X}^2$ and 
\begin{eqnarray}
J^\alpha=2\mathcal{X}\overline{E}m^\alpha + 2\mathcal{X}E\overline{m}^\alpha.
\end{eqnarray}
Using $B\overline{B} = 2\mathcal{X}^2$ and $\left(E\left(2\mathcal{X}^2 + B\overline{B}\right)+B\right) = 0$, we can write: $E = \frac{1}{2\sqrt{2}\mathcal{X}}e^{i\psi}$  and $B = -\sqrt{2}\mathcal{X}e^{i\psi}$, where $\psi$ is a real function of spacetime coordinates.
Given the imposed constraints arising from spherical symmetry, we can now express the null bases of the null frame $(l^\alpha, p^\alpha, M^\alpha, \overline{M}^\alpha)$ as:
\begin{widetext}
\begin{eqnarray}
    l^\mu &=& A~k^\mu_+ + \frac{1}{8~\mathcal{X}^2A} k^\mu_- + \frac{e^{-i\psi}}{2\sqrt2\mathcal{X} }m^\mu + \frac{e^{i\psi}}{2\sqrt2\mathcal{X} }\overline{m}^\mu,\nonumber\\
    p^\mu &=& 2\mathcal{X}^2A~k^\mu_+ + \frac{1}{4A} k^\mu_- - \frac{\mathcal{X}e^{-i\psi}}{\sqrt2}m^\mu - \frac{\mathcal{X}e^{i\psi}}{\sqrt2}\overline{m}^\mu,\nonumber\\
    M^\mu &=& e^{i\Theta}\left(-\sqrt{2}A\mathcal{X}~e^{i\psi} k^\mu_+ + \frac{e^{i\psi}}{4\sqrt2 A\mathcal{X}} k^\mu_- + \frac12 m^\mu - \frac{e^{2i\psi}}{2} \overline{m}^\mu\right),\nonumber\\
    \overline{M}^\mu &=& e^{-i\Theta}\left(-\sqrt{2}A\mathcal{X}~e^{-i\psi} k^\mu_+ + \frac{e^{-i\psi}}{4\sqrt2 A\mathcal{X}} k^\mu_- - \frac{e^{-2i\psi}}{2} m^\mu + \frac{e^{2i\psi}}{2} \overline{m}^\mu\right),
    \label{nullframe}
\end{eqnarray}
\end{widetext}
which implies a composite of Lorentz transformations, comprising a boost in the $(k^\alpha_+, k^\alpha_-)$ plane with a boost parameter of $A$, followed by a null rotation around $k^\alpha_-$ with $E = \frac{1}{2\sqrt{2}\mathcal{X}}e^{i\psi}$, and another null rotation around $k^\alpha_+$ with $B = -\sqrt{2}\mathcal{X}e^{i\psi}$, and a spatial rotation of $\Theta$, converts the invariant null frame $(k^\alpha_+, k^\alpha_-, m^\alpha, \overline{m}^\alpha)$ to the null frame $(l^\alpha, p^\alpha, M^\alpha, \overline{M}^\alpha)$, where $l^\alpha\in T_p(S)$. Now, if the three-dimensional hypersurface $S$ is the photon surface then there would be further constraints on the boost parameter $A$ and the null rotation parameter $E$.

One can verify that, for the spherically symmetric solution of Einstein's equations, the following conditions consistently hold:
\begin{align}
    \pi &= \kappa = \nu = \tau = \lambda = \sigma = 0, \nonumber \\
    \alpha + \overline{\beta} &= 0.
\end{align}
Consequently, for solutions of this nature, the previously mentioned photon surface condition (Eq.~(\ref{PScon1})) undergoes simplification to:
\begin{widetext}
    \begin{eqnarray}
        \left(\frac{E\overline{E}}{A}\left(\mu + \overline{\mu}\right) + (\rho + \overline{\rho})A\right) + 2\left(\frac{E\overline{E}}{A}\left(\gamma + \overline{\gamma}\right) + (\epsilon + \overline{\epsilon})A\right)+\frac{2}{A}\left(A~DA +\frac{E\overline{E}}{A}~\Delta A\right) + \left(\overline{E}~DE+E~D\overline{E}+\frac{\Delta E}{E}+\frac{\Delta \overline{E}}{\overline{E}}\right)= 0\,\, , \nonumber\\ \label{PScon3}  \end{eqnarray}
    \end{widetext}
where we use the Eq.~(\ref{PScon1}) and it's complex conjugate counterpart. For the spherically symmetric scenario, since the $E\propto e^{i\psi}$, the last term written in the round bracket $\left(\overline{E}~DE+E~D\overline{E}+\frac{\Delta E}{E}+\frac{\Delta \overline{E}}{\overline{E}}\right)$ is identically zero.
\subsection{Photon surfaces in spherically symmetric, static, Petrov type-D solution}
On top of spherical symmetry, if we consider a static solution (i.e., $SO(3)\times\mathbb{R}$ symmetry) then the above equation simplifies further to:
\begin{eqnarray}
    \rho +2\epsilon = 0,
    \label{PScon4}
\end{eqnarray}
where for the static case the boost parameter $A$ becomes one (i.e., zero boost) and $\mu = \rho$ and $\gamma = \epsilon$. From $\Tilde{\kappa}=0$, one can show that $A = 1$ for a spherically symmetric static case. Now, using Eq.~(\ref{PScon2}) we can write:
\begin{eqnarray}
\Phi_{00}-8\epsilon^2-\Psi_2 -\frac{R}{12} = 0\, ,
\label{PScon5}
\end{eqnarray}
where we consider a Petrov type-D solution. Here, $\Phi_{00}=\frac12 R_{\mu\nu}l^\mu l^\nu$ and $R$ is Ricci scalar. In a spherically symmetric, type-D, static solution, it can be confirmed that if the condition mentioned above is satisfied at a specific radial distance from the center, then a photon sphere of that radius must exist. 

To prove that we need to consider a general, static, spherically symmetric spacetime as follows:
\begin{eqnarray}
    dS^2 = -\mathcal{A}(r)dt^2+\mathcal{B}(r) dr^2 + r^2 \mathcal{C}(r)(d\theta^2 + \sin^2\theta d\phi^2)\, ,\nonumber\\
\end{eqnarray}
where $\mathcal{A}(r), \mathcal{B}(r)$ and $\mathcal{C}(r)$ are the positive definite functions of $r$. Now one can construct a null frame at any point of the above spacetime consisting of the following null bases:
\begin{eqnarray}
    k^\mu_+ &=& \left\{\frac{1}{\sqrt{2\mathcal{A}(r)}}, \frac{1}{\sqrt{2\mathcal{B}(r)}}, 0, 0\right\},\nonumber\\
    k^\mu_- &=& \left\{\frac{1}{\sqrt{2\mathcal{A}(r)}}, -\frac{1}{\sqrt{2\mathcal{B}(r)}}, 0, 0\right\},\nonumber\\
    m^\mu &=& \left\{0, 0, \frac{1}{r\sqrt{2\mathcal{C}(r)}}, \frac{i}{r\sin\theta\sqrt{2\mathcal{C}(r)}}\right\},\nonumber\\
    \overline{m}^\mu &=& \left\{0, 0, \frac{1}{r\sqrt{2\mathcal{C}(r)}}, \frac{-i}{r\sin\theta\sqrt{2\mathcal{C}(r)}}\right\}.
    \label{nullframe1}
\end{eqnarray}
Now in this null frame, the components of the Weyl curvature tensor have a canonical form since only the non-zero component is $\Psi_2:$
\begin{widetext}
\begin{eqnarray}
    \Psi_2 &=& \frac{\left(-r^2 \mathcal{B}(r) \mathcal{C}^2(r) \mathcal{A}'(r)^2 + r \mathcal{A}(r) \mathcal{C}(r) \left(-\mathcal{A}'(r) \left(r \mathcal{C}(r) \mathcal{B}'(r) + \mathcal{B}(r) \left(2 \mathcal{C}(r) + r \mathcal{C}'(r)\right)\right) + 2 r \mathcal{B}(r) \mathcal{C}(r) \mathcal{A}''(r)\right)\right)}{24~ r^2 \mathcal{A}^2(r) \mathcal{B}^2(r) \mathcal{C}^2(r)}\nonumber\\ &+& \frac{ \left(-4 \mathcal{B}(r)^2 \mathcal{C}(r) + r \mathcal{C}(r) \mathcal{B}'(r) \left(2 \mathcal{C}(r) + r \mathcal{C}'(r)\right) + 2 \mathcal{B}(r) \left(2 \mathcal{C}(r)^2 + r^2 \left(\mathcal{C}'\right)^2 - r^2 \mathcal{C}(r) \mathcal{C}''(r)\right)\right)}{24~ r^2  \mathcal{B}^2(r) \mathcal{C}^2(r)}\, ,
\end{eqnarray}
and the non-zero components of the Ricci tensor are:
\begin{eqnarray}
    \Phi_{00} &=& \Phi_{22} = \frac{\left(2 \mathcal{C}^2(r) \mathcal{A}'(r) + r \mathcal{A}(r) \left(\mathcal{C}'(r)\right)^2 +\mathcal{C}(r) \left((-4 \mathcal{A}(r) + r \mathcal{A}'(r)) \mathcal{C}'(r) - 2 r \mathcal{A}(r) \mathcal{C}''(r)\right)\right)}{8 r \mathcal{A}(r) \mathcal{B}(r) \mathcal{C}^2(r)}+\frac{ \mathcal{B}'(r) \left(2 C1(r) + r \mathcal{C}'(r)\right) }{8 r \mathcal{B}(r)^2 \mathcal{C}(r)}\, ,\nonumber\\
    \\
    \Phi_{11} &=& \frac{-r^2 \mathcal{B}(r) \mathcal{C}^2(r) \mathcal{A}'(r)^2 + \mathcal{A}^2(r) \mathcal{B}(r) \left(4 \mathcal{B}(r) \mathcal{C}(r) - \left(2 \mathcal{C}(r) + r \mathcal{C}'(r)\right)^2\right) + r^2 \mathcal{A}(r) \mathcal{C}^2(r) \left(-\mathcal{A}'(r) \mathcal{B}'(r) + 2 \mathcal{B}(r) \mathcal{A}''(r)\right)}{16 r^2 \mathcal{A}^2(r) \mathcal{B}^2(r) \mathcal{C}^2(r)}\, ,\nonumber\\
\end{eqnarray}
and the Ricci scalar is:
\begin{eqnarray}
    R &=& \frac{r^2 \mathcal{B}(r) \mathcal{C}(r)^2 \left(\mathcal{A}'(r)\right)^2 + r \mathcal{A}(r) \mathcal{C}(r) \left(r \mathcal{C}(r) \mathcal{A}'(r) \mathcal{B}'(r) - 2 \mathcal{B}(r) \left(\mathcal{A}'(r) \left(2 \mathcal{C}(r) + r \mathcal{C}'(r)\right) + r \mathcal{C}(r) \mathcal{A}''(r)\right)\right)}{2 r^2 \mathcal{A}(r)^2 \mathcal{B}(r)^2 \mathcal{C}(r)^2}\nonumber\\ &+& \frac{\left(4 \mathcal{B}(r)^2 \mathcal{C}(r) + 2 r \mathcal{C}(r) \mathcal{B}'(r) \left(2 \mathcal{C}(r) + r \mathcal{C}'(r)\right) + \mathcal{B}(r) \left(-4 \mathcal{C}(r)^2 + r^2 \left(\mathcal{C}'(r)\right)^2 - 4 r \mathcal{C}(r) \left(3 \mathcal{C}'(r) + r \mathcal{C}''(r)\right)\right)\right)}{2 r^2 \mathcal{B}(r)^2 \mathcal{C}(r)^2}\, .\nonumber\\
\end{eqnarray}
\end{widetext}
From the above expressions of the non-zero components of the curvature tensor, it can be understood that the null frame described in Eq.~(\ref{nullframe1}) is indeed a canonical frame. Now, we need to use the CK algorithm to fix the frame further. 
As we have observed above, at the zeroth order (i.e., $q=0$), the components of the Riemann curvature tensor have the canonical form where $\Psi_2, \Phi_{00}, \Phi_{11}$ and $\Lambda$ (i.e., $\frac{R}{24}$) are four algebraically independent components of the Riemann tensor and all of these components are functions of $r$ only. Therefore, $t_0 =1$. Now, since under boost and a null rotation the canonical form of the components of the Riemann curvature tensor would not be preserved, the isotropy group at the zeroth order has only spatial spins; i.e, $\hat{I}_0 =1$. Therefore, at the zeroth order, by using the CK algorithm, we fix the frame $(k^\alpha_+, k^\alpha_-, m^\alpha,\overline{m}^\alpha)$ described in Eq.~(\ref{nullframe1}) up to spatial spins.  Given the static and spherically symmetric nature of the spacetime manifold, it can be demonstrated that at the first order, both $t_1$ and $\hat{I}_1$ retain their values from the zeroth order. Therefore, the algorithm stops here and as we stated before, in the spherically symmetric scenario, we cannot completely fix the frame. Hence, the frame $(k^\alpha_+, k^\alpha_-, m^\alpha,\overline{m}^\alpha)$ is invariantly defined up to the spatial spin.  The complete set of algebraically independent Cartan scalars are:
\begin{eqnarray}
    \Psi_2, \Phi_{00}, \Phi_{11}, \Lambda, \epsilon, \rho, \text{\it{and their frame derivatives}}.
\end{eqnarray}

Hence, the condition for the photon surface expressed in Eq.~(\ref{PScon5}) represents a specific relation among the Cartan scalars defined in the invariant null frame (\ref{nullframe1}), and this relation holds exclusively on the photon surface. Using Eq.~(\ref{PScon5}) and the above expressions of $\Psi_2, R, \Phi_{00}$ we get:
\begin{eqnarray}
   &&\frac{\mathcal{A}'(-r \mathcal{C}(r) \mathcal{A}' + \mathcal{A}(r) \left(2 \mathcal{C}(r) + r \mathcal{C}'(r)\right))}{4 r \mathcal{A}(r)^2 \mathcal{B}(r) \mathcal{C}(r)}\Big|_{r=r_{PS}} = 0\,\, ,\nonumber\\
&\implies& r_{PS} = \frac{\mathcal{A}(r) \left(2 \mathcal{C}(r) + r \mathcal{C}'(r)\right))}{\mathcal{C}(r) \mathcal{A}'(r)}\Big|_{r=r_{PS}} \, ,   
\end{eqnarray}
where $r_{PS}$ is the radius of photon surface and $\epsilon = \frac{\mathcal{A}'}{4 \sqrt{2} \mathcal{A}(r) \sqrt{\mathcal{B}(r)}}$. We can determine the radius of the photon surface, if it exists in a given spacetime manifold, by solving the aforementioned algebraic equation for $r_{PS}$. For Schwarzschild spacetime, the solution of the above equation is $r_{PS} = 3M$, where $M$ is the Schwarzschild mass.

The photon surface condition: $\Phi_{00}-8\epsilon^2-\Psi_2 -\frac{R}{12} = 0$ gives us a novel insight understanding how the presence of matter influences the existence of photon surfaces. In the vacuum, the condition becomes $\Psi_2 = -8\epsilon^2$.

\begin{figure*}\label{GammavariationQ}
{\includegraphics[width=70mm,height=96mm]{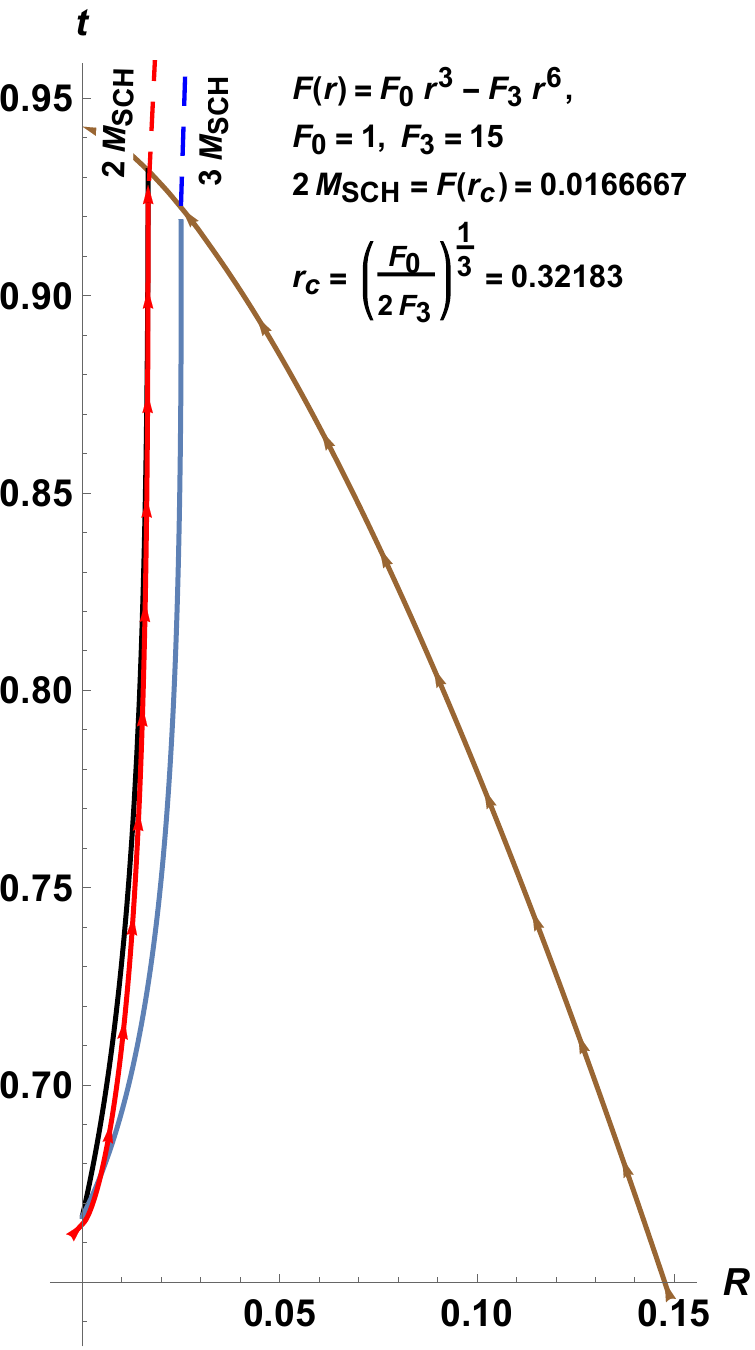}}
\hspace{0.2cm}
{\includegraphics[width=70mm,height=96mm]{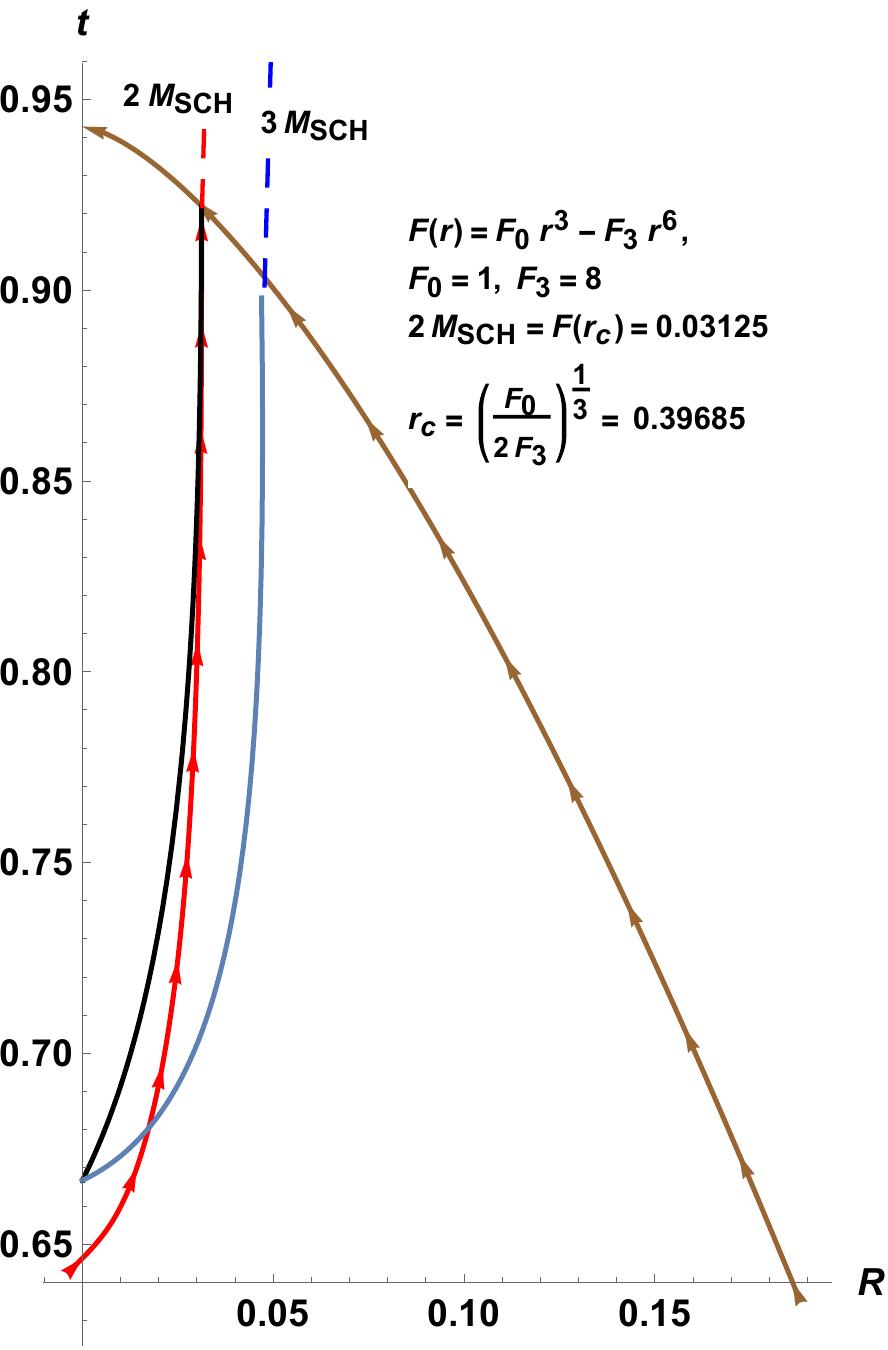}}
\hspace{0.2cm}
{\includegraphics[width=70mm,height=96mm]{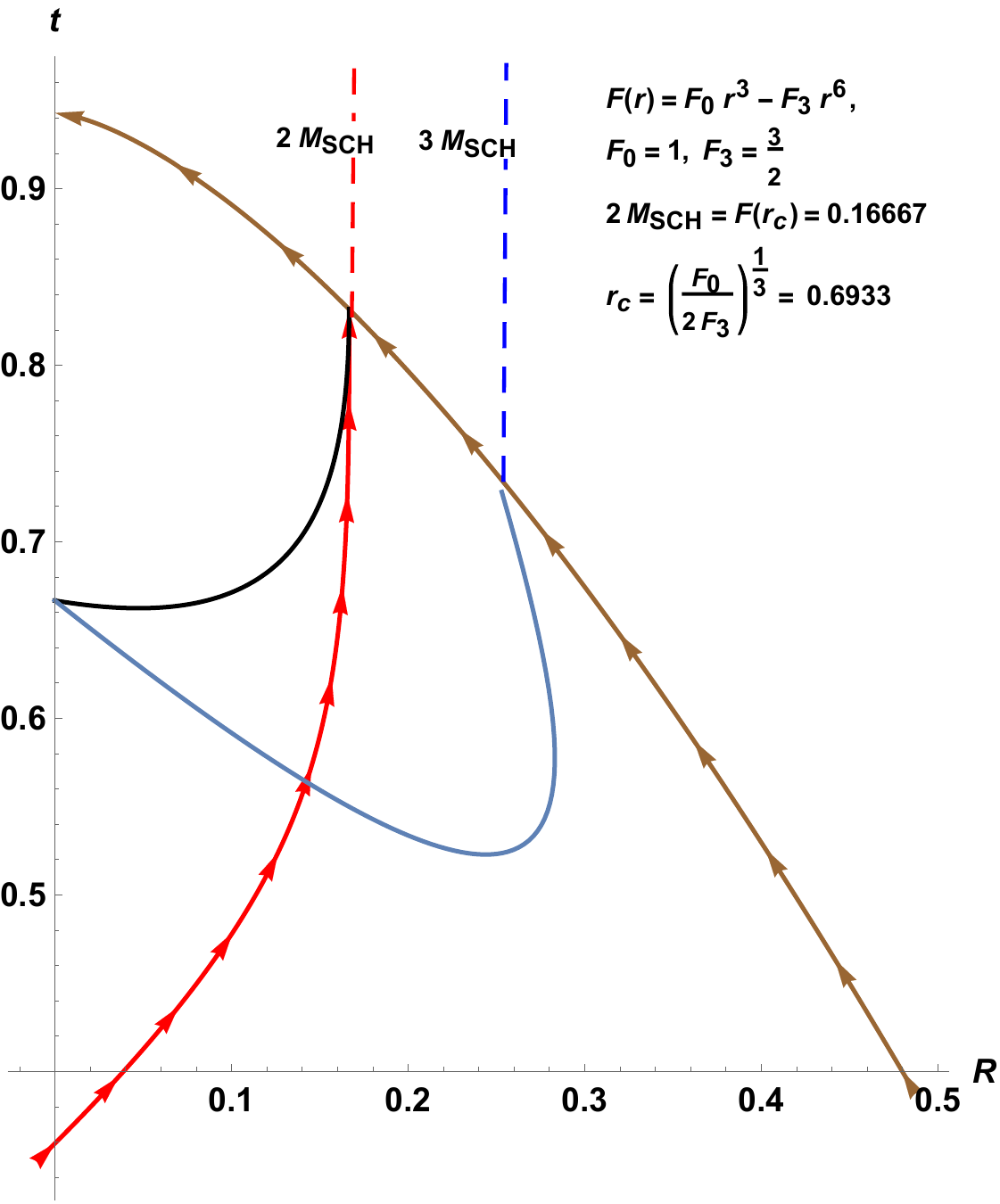}}
\hspace{0.2cm}
{\includegraphics[width=70mm,height=96mm]{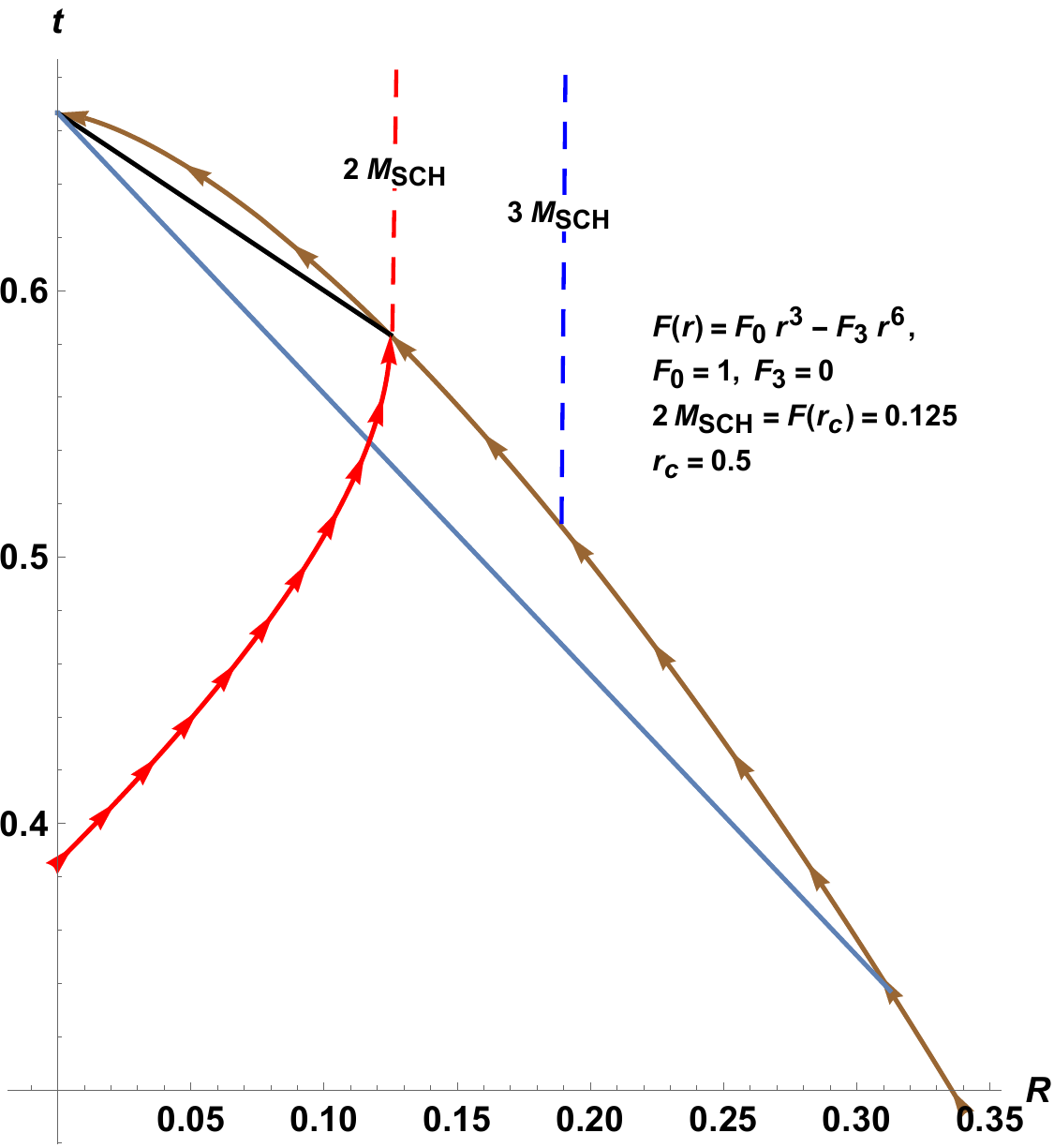}}
\caption{Figure depicts the evolution of the photon surfaces in LTB collapse for various values of $F_3$, where the boundary of collapsing cloud, apparent horizon, and photon surface are shown by the solid brown, black, and blue curves, respectively.}
\label{Psfigure}
\end{figure*}
\subsection{Photon surfaces in marginally bound LTB solutions:} 
The line element of the Lemaitre-Tolman-Bondi (LTB) spacetime can be written as:
\begin{eqnarray}
    dS^2 = -dt^2 + \frac{\mathcal{R}^{\prime 2}(t,r)}{1+E(r)}~dr^2 +\mathcal{R}^2(t,r)~d\Omega
\end{eqnarray}
where $\mathcal{R}(t,r)$ is the physical radius and $E(r)$ is the velocity function. For the simplest scenario, here we consider marginally bound LTB spacetime where $E(r)=0$. We can construct the following null frame: 
\begin{eqnarray}
    k^\mu_+ &=& \left\{\frac{1}{\sqrt{2}}, \frac{1}{\sqrt{2}~\mathcal{R}^\prime(t,r)}, 0, 0\right\},\nonumber\\
    k^\mu_- &=& \left\{\frac{1}{\sqrt{2}}, -\frac{1}{\sqrt{2}~\mathcal{R}^\prime(t,r)}, 0, 0\right\},\nonumber\\
    m^\mu &=& \left\{0, 0, \frac{1}{\sqrt{2}~\mathcal{R}(t,r)}, \frac{i}{\sqrt{2}~\mathcal{R}(t,r)~\sin\theta}\right\},\nonumber\\
    \overline{m}^\mu &=& \left\{0, 0, \frac{1}{\sqrt{2}~\mathcal{R}(t,r)}, \frac{-i}{\sqrt{2}~\mathcal{R}(t,r)~\sin\theta}\right\}.
    \label{nullframe2}
\end{eqnarray}
In this null frame, the following correspond to the non-zero components of the Riemann curvature tensor:
    \begin{eqnarray}
    \Psi_2 &=& \frac{-\dot{\mathcal{R}}^2 + \mathcal{R}\Ddot{R}}{6\mathcal{R}^2}+\frac{\dot{\mathcal{R}}\dot{\mathcal{R}}^\prime - \mathcal{R}\Ddot{\mathcal{R}}^\prime}{6\mathcal{R}\mathcal{R}^\prime}\, ,\\
    \Phi_{00} &=& \Phi_{22} =\frac{\dot{\mathcal{R}}\dot{\mathcal{R}}^\prime - \mathcal{R}^\prime\Ddot{\mathcal{R}}}{2\mathcal{R}\mathcal{R}^\prime}\, ,
    \end{eqnarray}
    \begin{eqnarray}
    \Phi_{11} &=& \frac14\left(\frac{\dot{\mathcal{R}}^2}{\mathcal{R}^2}-\frac{\Ddot{\mathcal{R}}^\prime}{\mathcal{R}^\prime}\right)\, ,\\
    R &=& \frac{2\left(\dot{\mathcal{R}}^2+2\mathcal{R}\Ddot{\mathcal{R}}\right)}{\mathcal{R}^2}+\frac{2\left(2\dot{\mathcal{R}}\dot{\mathcal{R}}^\prime+\mathcal{R}\Ddot{\mathcal{R}}^\prime\right)}{\mathcal{R}\mathcal{R}^\prime}.
\end{eqnarray}
Therefore, in the null frame constructed above (Eq.~(\ref{nullframe2})), the spin frame components of the Riemann curvature tensor take the canonical form (i.e., Petrov type-D). Consequently, similar to the static case, here we can also completely fix the above frame by using the CK algorithm. Using the expression of the null bases corresponding to the invariant frame $(k^\alpha_+, k^\alpha_-, m^\alpha,\overline{m}^\alpha)$ (Eq.~(\ref{nullframe2})) and the non-canonical null frame $(l^\alpha, p^\alpha, M^\alpha, \overline{M}^\alpha)$ defined in Eq.~(\ref{nullframe}), we can write:
\begin{widetext}
\begin{eqnarray}
    l^\mu &=& \left\{\frac{8\mathcal{X}^2A(t,r)^2+1}{8 \sqrt{2}~A(t,r)\mathcal{X}^2},~\frac{8\mathcal{X}^2A(t,r)^2-1}{8 \sqrt{2}~A(t,r)~\mathcal{X}^2\mathcal{R}^\prime},~\frac{\cos{(\psi(t,r))}}{2\mathcal{X} \mathcal{R}},~\frac{\sin{(\psi(t,r))}}{2\mathcal{X} \sin\theta~ \mathcal{R}}\right\}\, ,\nonumber\\
    p^\mu &=&\left\{\frac{8\mathcal{X}^2A(t,r)^2+1}{4\sqrt{2}~A(t,r)},~\frac{8\mathcal{X}^2A(t,r)^2-1}{4 \sqrt{2}~A(t,r)~\mathcal{R}^\prime},~\frac{-\cos{(\psi(t,r))}\mathcal{X}}{ \mathcal{R}},~\frac{-\sin{(\psi(t,r))}\mathcal{X}}{ \sin\theta~ \mathcal{R}}\right\}\, ,\nonumber\\
    M^\mu &=& \left\{\frac{e^{i \psi (t,r)} \left(1-8 \mathcal{X}^2 A(t,r)^2\right)}{8 \mathcal{X} A(t,r)},-\frac{e^{i \psi (t,r)} \left(8 \mathcal{X}^2 A(t,r)^2+1\right)}{8 \mathcal{X} A(t,r) \mathcal{R}^\prime},-\frac{-1+e^{2 i \psi (t,r)}}{2 \sqrt{2}~\mathcal{R}},\frac{i \left(1+e^{2 i \psi (t,r)}\right)}{2 \sqrt{2}~\mathcal{R}\sin\theta}\right\} ,\nonumber\\
    \overline{M}^\mu &=& \left\{\frac{e^{i \psi (t,r)} \left(1-8 \mathcal{X}^2 A(t,r)^2\right)}{8 \mathcal{X} A(t,r)},-\frac{e^{i \psi (t,r)} \left(8 \mathcal{X}^2 A(t,r)^2+1\right)}{8 \mathcal{X} A(t,r) \mathcal{R}^\prime},-\frac{-1+e^{-2 i \psi (t,r)}}{2 \sqrt{2}~\mathcal{R}},-\frac{i \left(1+e^{-2 i \psi (t,r)}\right)}{2 \sqrt{2}~\mathcal{R}\sin\theta}\right\}\, .\nonumber\\
\end{eqnarray}
\end{widetext}

Now, since $l^\mu$ should satisfy the geodesic equation, $\Tilde{D}l_0 = \Tilde{D}l_1 = \Tilde{D}l_2 = \Tilde{D}l_3 = 0$; i.e., $\Tilde{\kappa} =0$ and $\Tilde{\epsilon} + \overline{\Tilde{\epsilon}} =0$. From the azimuthal part of the geodesic equation, we get the following expression for $A(t,r)$ with the extra constraint $\theta = \frac\pi2$:
\begin{eqnarray}
    A(t,r) = \sqrt{\frac{1-\dot{\mathcal{R}}}{8\mathcal{X}^2\left(1+\dot{\mathcal{R}}\right)}}.
\end{eqnarray}
\begin{figure*}
{\includegraphics[width=85mm,height=95mm]{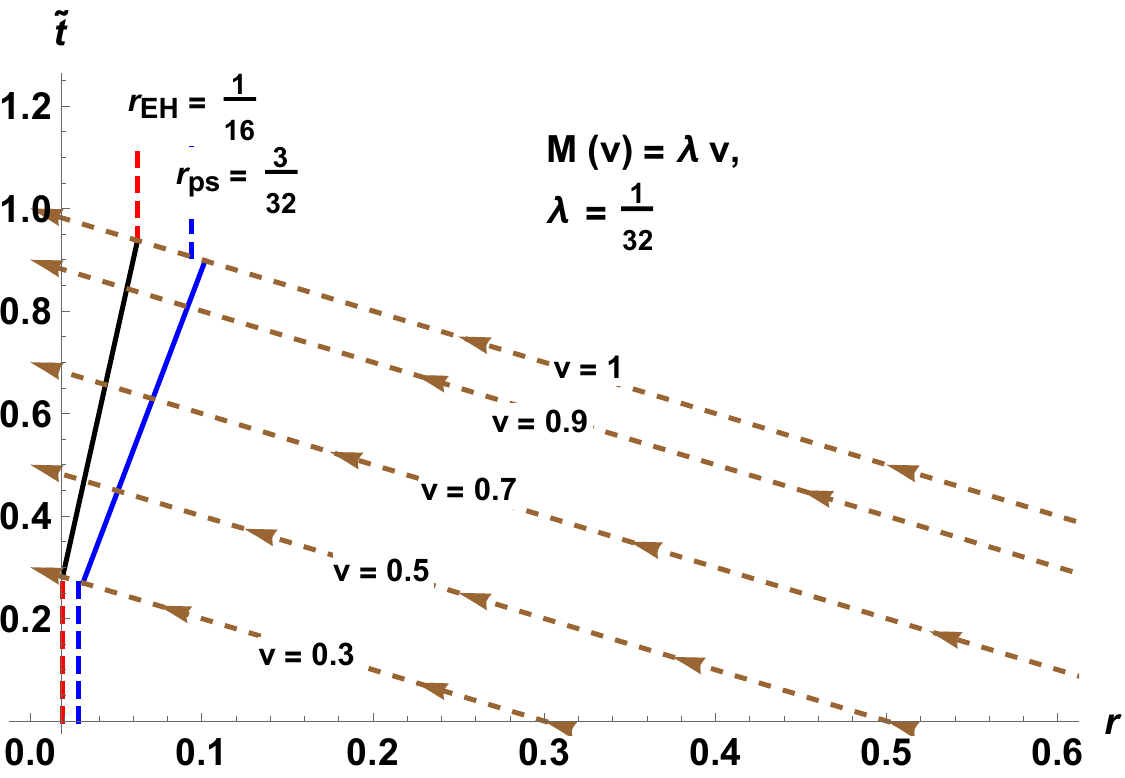}}
\hspace{0.5cm}
{\includegraphics[width=85mm,height=95mm]{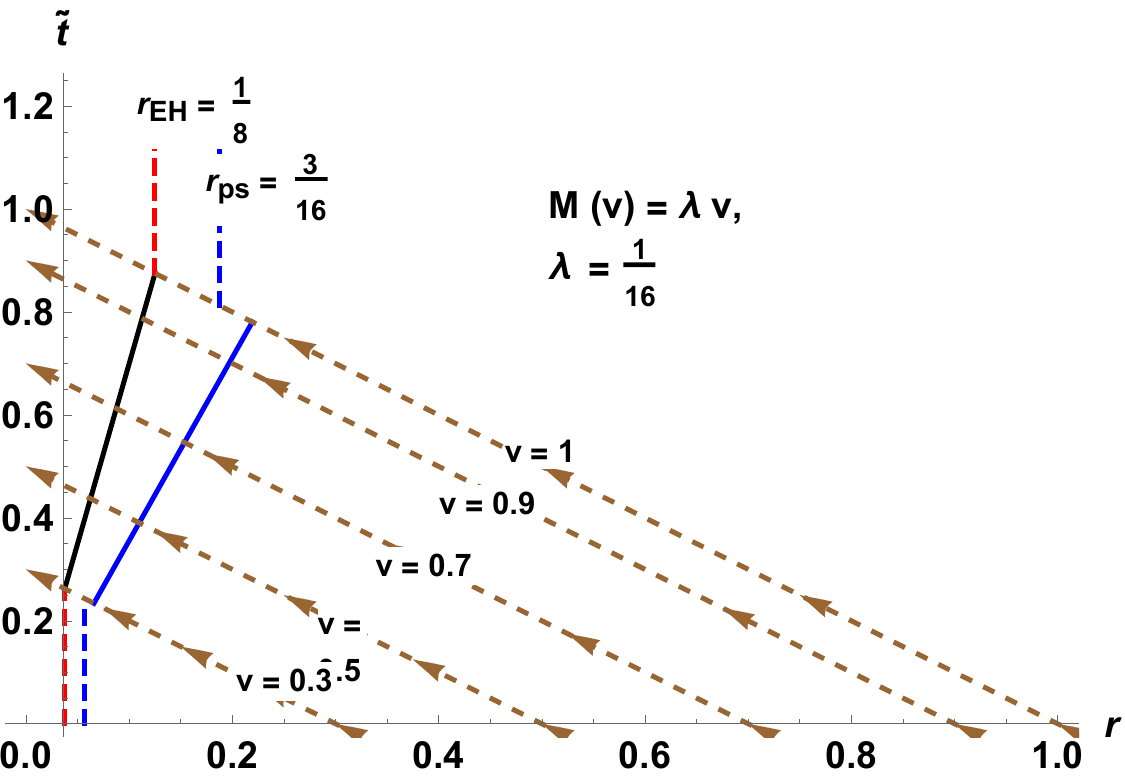}}
\caption{Figure illustrates the dynamics of photon surfaces (highlighted in blue) and apparent horizons (highlighted in black) in Vaidya spacetime in the $(\Tilde{t}, r)$ frame, where $\Tilde{t} = v -r$. The brown dotted arrow lines depict the inflow of matter in the imploding Vaidya spacetime, where we consider $M(v) = \lambda v$. In the first figure (from the left) and second figure we consider $\lambda = \frac{1}{32}$ and $\lambda = \frac{1}{16}$, respectively. }
\label{Psfigure2}
\end{figure*}
Now, using the above expression for $A(t,r)$, one can show $\Tilde{D}l_0 = \Tilde{D}l_1 = 0$, which implies the following condition: 
    \begin{eqnarray}
        \mathcal{R} \dot{\mathcal{R}}^3 \dot{\mathcal{R}}^\prime-\mathcal{R}\mathcal{R}^\prime \Ddot{\mathcal{R}}-\mathcal{R}^\prime\left(\dot{\mathcal{R}}^2-1\right)^2=0.
        \label{PScon6}
    \end{eqnarray}
In Eq.~(\ref{PScon3}), we found the photon surface condition defined in the invariant null frame $(k^\alpha_+, k^\alpha_-, m^\alpha,\overline{m}^\alpha)$. The above condition can be derived from that previous condition using the above expression of $A(t,r)$. Therefore, the condition given in Eq.~(\ref{PScon6}) also holds on the photon surface $S$. The solution of the physical radius $\mathcal{R}(t,r)$ can be derived from the solution of the Einstein equations, and therefore, Eq.~(\ref{PScon6}) is an algebraic equation in terms of $t,r$. For the marginally bound LTB spacetime, $\mathcal{R}(t,r)$ has the following form:
\begin{eqnarray}
    \mathcal{R}(t,r)=\left(\frac{3}{2}\sqrt{F(r)}\left(t_s(r)-t\right)\right)^{2/3}\, ,
\end{eqnarray}
where $F(r)$ is the Misner-Sharp mass and the $t_s(r) = \frac23 \frac{r^\frac32}{\sqrt{F(r)}}$ is the singularity formation time. To model an inhomogeneous gravitational collapse, we consider $F(r) = F_0 r^3 - F_3 r^6$, where $F_0> 0$ and $F_3 >0$. The evolution of photon surfaces in LTB spacetime is depicted in Fig.~(\ref{Psfigure}), and it can be observed that the evolution of the photon surfaces does not depend upon the global picture of the spacetime manifold. Here, we consider a spacetime structure which is internally LTB and externally Schwarzschild spacetime. In this paper, we define the photon surface in the spin frame which is purely a local definition and, therefore, any change in the external spacetime does not influence the evolution of the photon surface in the LTB interior. Therefore, for a homogeneous scenario depicted in the fourth figure in Fig.~(\ref{Psfigure}), the internal photon surface does not match the external one at the boundary of the collapsing cloud. In other figures, it can be seen that the external and internal photon surfaces merge at the boundary because of the null value of the density at the boundary of the collapsing cloud. Our analysis introduces a novel perspective on the causal structure of the singularity. While previous research established that a radial null geodesic can escape from the central singularity when $\frac{F_3}{F_0^\frac52} > 25.99$ \cite{Dwivedi1, Dwivedi2, Singh1, Singh2, Goswami1}, our findings indicate that considering circulating null geodesics on an evolving photon surface relaxes this constraint to $\frac{F_3}{F_0^\frac52} > 8$. However, it should be noted that our analysis only shows the local causal structure of the singularity.

\subsection{Photon surfaces in Vaidya spacetimes}
The imploding Vaidya spacetime is a collapsing spherically symmetric null-dust solution of Einstein's equations. The line element of this spacetime can be written as:
\begin{eqnarray}
    dS^2 = - \left(1-\frac{2M(v)}{r}\right) dv^2 + 2dr~dv + r^2 d\Omega\, ,
\end{eqnarray}
where $v$ is a null coordinate defined as: $v=t+r+2M(v)\log\left(\frac{r}{2M(v)}-1\right)$ and the ADM mass $M(v)$ is a positive definite function of $v$. One can construct the following null frame at any point in this spacetime manifold:
\begin{eqnarray}
    k^\mu_+ &=& \left\{1, \frac12\left(1-\frac{2M(v)}{r}\right), 0, 0\right\},\nonumber\\
    k^\mu_- &=& \left\{0, -1, 0, 0\right\},\nonumber\\
    m^\mu &=& \left\{0, 0, \frac{1}{\sqrt{2}r}, \frac{i}{\sqrt{2}r\sin\theta}\right\},\nonumber\\
    \overline{m}^\mu &=& \left\{0, 0, \frac{1}{\sqrt{2}r}, \frac{-i}{\sqrt{2}r\sin\theta}\right\}.
    \label{nullframe3}
\end{eqnarray}
In this null frame, the non-zero spin-frame components of the Riemann curvature tensor are:
\begin{eqnarray}
    \Psi_2 &=& -\frac{2M(v)}{r^3}\, ,\nonumber\\
    \Phi_{00} &=& \frac{M^\prime (v)}{r^2},
\end{eqnarray}
and the non-zero spin coefficients are:
\begin{eqnarray}
    \rho = -\frac{1}{2r}\left(1-\frac{2M(v)}{r}\right),\, \mu = -\frac1r,\,\epsilon = \frac{M(v)}{2r^2}.
\end{eqnarray}
It can be seen that the above expressions of the non-zero components of the curvature tensor have the Petrov type-D form in the frame $(k^\alpha_+, k^\alpha_-, m^\alpha,\overline{m}^\alpha)$ defined in Eq.~(\ref{nullframe3}). Therefore, like in the previous scenarios, here we can invariantly fix the frame up to spatial spin using the CK algorithm. The complete set of algebraically independent Cartan scalars would be: 
$$\Psi_2, \Phi_{00}, \rho, \epsilon, \text{and their frame derivatives}.$$
Now, using the condition for a photon surface written in Eq.~(\ref{PScon3}) and the above expressions for the spin coefficients, we can write:
\begin{eqnarray}
    -\left(1-\frac{3M(v)}{r_{PS}}\right)\left(1-\frac{2M(v)}{r_{PS}}\right)~+~M^\prime(v) = 0,
\end{eqnarray}
which implies $r_{PS} = 3M(v)$ is not the radius of the photon surface in an imploding Vaidya spacetime since there exists a non-zero contribution of the local flux of matter through $M^\prime (v)$ to the radius of the photon surface. Solving the above equation for $r_{PS}$, we get the following solution:
\begin{eqnarray}
 r_{PS}(v)~=~\frac{-5 M(v)+ M(v)\left(-\sqrt{24 M'(v)+1}\right)}{2 \left(M'(v)-1\right)}.
\end{eqnarray}
The above result implies that if $M^\prime(v)$ is discontinuous at a specific value of $v$ then the $r_{PS}$ would also be discontinuous at that value of $v$. 
Therefore, for $M(v)$ of the following type\\
\begin{eqnarray}M(v)= 
\begin{dcases}
    M(v)&: M^\prime(v) \neq 0 ~\text{if } 0\leq v\leq v_0\\
    M_0&: M^\prime(v) = 0, ~\text{if } v> v_0
\end{dcases}
\end{eqnarray}
the radius of the photon surface, $r_{PS}$, would not be continuous at $v_0$. In Fig.~(\ref{Psfigure2}), we show the dynamics of photon surfaces and apparent horizons in the imploding Vaidya spacetime with $M(v) = \lambda v$. It can be seen that after $v=1$, $M^\prime = 0$, the spacetime becomes a Schwarzschild spacetime with Schwarzschild mass $M_0 =\frac2\lambda$. Therefore, as stated above, the photon surface dynamics in general become discontinuous at the boundary $v=1$.
\section{Conclusion}
\label{sec5}
 In this paper, we establish an invariant definition of a photon surface within a spin frame. Initially, we derive the spin frame version of the tensorial photon surface conditions using spin coefficients. Subsequently, utilizing this condition, we derive the corresponding conditions expressed in terms of Cartan scalars defined within an invariant null frame. Here we consider only the Petrov type-D solutions and, therefore, we can invariantly identify two principal null directions. One can always find a null frame where the Riemann tensor and its derivatives take the type-D canonical form. Therefore, in that canonical frame, two of the null bases can be invariantly fixed using the Cartan-Karlhede algorithm. Subsequently, we then express the photon surface conditions in terms of spin coefficients of the null frame defined invariantly up to the spatial spin. To express the photon surface condition in terms of spin coefficients within the invariantly defined null frame, we identify the Lorentz transformation which converts the invariant null frame to a non-canonical frame, wherein one of the null bases becomes tangent to a three-dimensional hypersurface $S$. Importantly, the basis must adhere to the photon surface condition if and only if the hypersurface $S$ indeed qualifies as a photon surface. Since the spinor definition of the photon surfaces is completely local, the dynamics of the photon surface are independent of the global structure of the spacetime manifold. Utilizing the spinorial photon surface definition in different type-D solutions, we get the following important results:
 \begin{itemize}
     \item In static, spherically symmetric scenarios, we demonstrate the alignment of our findings with the established photon sphere conditions. Furthermore, when expressed in terms of Cartan scalars, the photon sphere condition reveals the impact of the presence of matter on the characteristics of the photon sphere. 
     \item We employ the spinor definition of photon surfaces within the marginally bound Lemaitre-Tolman-Bondi (LTB) and the imploding Vaidya spacetimes. In both scenarios, we investigate the evolution of photon surfaces within the collapsing dynamics. Our analysis reveals that in the LTB spacetime, there are parameter spaces where photon surfaces exhibit expansion, originating from the central singularity. Consequently, null geodesics confined to these surfaces can initiate their trajectory in close proximity to the central singularity region and, therefore, these geodesics have the potential to carry information about any new physics that may emerge in the ultra-high density regions near the singularity. This phenomenon could lead to novel observational consequences with significant implications.
 \end{itemize}
 A comparable spin-frame analysis can be conducted for rotating spacetimes, as well as for collapsing non-spherical Szekeres space-times, to explore the dynamics of photon surfaces in these scenarios. However, we defer this exploration to future work.

 \section{Acknowledgement}

DD would like to acknowledge the support of
the Atlantic Association for Research in the Mathematical Sciences (AARMS) for funding the work. AC acknowledges financial support from NSERC.

\end{document}